\documentstyle[12pt,A4,epsf]{article}
\begin{document}
\hyphenation{ef-fec-tive}
\hyphenation{ac-tion}
\hyphenation{quan-tum}
\def\be{\begin{equation}}
\def\ee{\end{equation}}
\def\bea{\begin{eqnarray}}
\def\eea{\end{eqnarray}}
\def\nn{\nonumber \\}
\def\cF{{\cal F}}
\def\det{{\rm det\,}}
\def\Tr{{\rm Tr\,}}
\def\e{{\rm e}}
\def\etal{{\it et al.}}
\def\erp2{{\rm e}^{2\rho}}
\def\erm2{{\rm e}^{-2\rho}}
\def\er4{{\rm e}^{4\rho}}
\def\etal{{\it et al.}}

\title{Induced wormholes due to quantum effects of spherically reduced
matter in large $N$ approximation}
\author{
S. Nojiri\thanks{E-mail: nojiri@cc.nda.ac.jp}\\
{\it Department of Mathematics and Physics} \\
{\it National Defence Academy,
Hashirimizu Yokosuka 239, JAPAN}
\\[10pt]
O. Obregon\thanks{E-mail: octavio@ifug3.ugto.mx}
\\{\it Instituto de Fisica de la Universidad de Guanajuato}
\\{\it P.O. Box E-143, 37150 Leon Gto., Mexico}
\\[10pt]
S.D. Odintsov\thanks{E-mail: odintsov@mail.tomsknet.ru}
\ and K.E. Osetrin\thanks{E-mail: osetrin@tspu.edu.ru} \\
{\it Tomsk State Pedagogical University}\\
{\it 634041 Tomsk, RUSSIA}}
\date{}
\maketitle
\begin{abstract}
Using one-loop effective action in large $N$ and $s$-wave approximation we
discuss the possibility to induce primordial wormholes at the early
Universe. An analytical solution is found for self-consistent
primordial wormhole with constant radius. Numerical study gives the
wormhole solution with increasing throat radius and
increasing red-shift function.
 There is also some indication to the
possibility of a topological phase transition.
\end{abstract}

\newpage

It is  a well  known fact that spherical reduction of Einstein gravity
(see, for example \cite{8}) leads to specific 2d dilatonic gravity
which belongs to the general class of 2d dilatonic gravities \cite{9}.
Spherical reduction of 4d matter leads to 2d dilaton coupled matter
theories.

Recently, 2d conformal anomaly for dilaton coupled scalars has been
found in ref.\cite{1} (see also \cite{2,3,4}).
Integrating this conformal anomaly one finds the anomaly
induced effective action \cite{2,3} (in $s$-wave and large $N$
approximation) which was written in ref.\cite{6} in most complete form.
A wide spectrum of physical problems may be addressed using the above
effective action. Let us only mention some of most interesting:
 anti-evaporation of multiple horizon black holes \cite{6,10}.
Hawking radiation \cite{13} in dilatonic supergravity \cite{3},
quantum cosmological models in dilatonic supergravity
\cite{11}, study of semi-classical energy-momentum tensor
in the presence of dilaton \cite{12}, etc.

In the present work we would like to discuss one more of these physical
phenomena: the appearance of spherically symmetric wormholes due to
the anomaly
induced effective action. The wormholes represent handles of
topological origin. They may be considered as bridges joining two
different Universes or two separate regions of the same Universe
\cite{14}. It is known that wormholes cannot occur as solution of
classical gravity-matter theory due to violations of energy conditions
in classical relativity \cite{14}. Hence, it is natural to expect their
manifestations only at the quantum level (or yet in the theory like
Born-Infeld D-brane or M-theory, for review, see \cite{15}).

Recently, using quantum scalar stress-energy tensor calculated on
spherically symmetric background in \cite{16} the back reaction problem
(i.e. self-consistent wormholes in semiclassical gravity) has been
investigated in ref.\cite{17} (see also \cite{18}). It has been indeed
numerically found a semiclassical quantum solution representing
a wormhole connecting two asymptotically spatially flat regions \cite{17}.
This result is extremely interesting and it should be verified in another
models and (or) approximations, taking into account the fact that the
approach developed in \cite{17} is not completely free from some
serious drawbacks \cite{18}.

In the present work we show the possibility of inducing wormholes in  the
early Universe, making use of the effective action method \cite{5} (large $N$
and $s$-wave approximation is used).

We will start from the action of Einstein gravity with
$N$ minimal scalars
\be \label{e1}
S=-\frac{1}{16\pi G}\int
d^4x\,\sqrt{-g_{(4)}} \left(R^{(4)}-2\Lambda\right) +\frac{1}{2}
\sum_{i=1}^N\int
d^4x\,\sqrt{-g_{(4)}}\, g^{\alpha\beta}_{(4)}
\partial_\alpha \chi_i
\partial_\beta \chi_i,
\ee
where $\chi_i$ are scalars, $N$ is the number of scalars (in order to
apply the large $N$ approach, $N$ is considered to be large, $N\gg 1$), $G$
and $\Lambda$ are the  gravitational and cosmological constants.

The convenient choice for the spherically symmetric spacetime is the
following one
\be
\label{e2}
ds^2=g_{\mu\nu}dx^\mu dx^\nu+e^{-2\phi}d\Omega,
\ee
where  $\mu,\nu=0,1$, $g_{\mu\nu}$ and $\phi$ depend only on  $x^0$,
$x^1$ and $d\Omega$ corresponds to the two-dimensional sphere.

The action (\ref{e1}), reduced according to (\ref{e2}) takes the form
\be
\label{e3}
S_{red}=\int d^2x \sqrt{-g}\e^{-2\phi}
\left[-{1 \over 16\pi G}
\{R + 2(\nabla  \phi)^2 -2\Lambda + 2\e^{2\phi}\}
+ {1 \over 2}\sum_{i=1}^N(\nabla \chi_i)^2 \right].
\ee
Working in large $N$ and $s$-wave approximation, one can calculate the
quantum correction to $S_{red}$ (effective action). Using 2d conformal
anomaly for dilaton coupled scalar, calculated in \cite{1} (see also
\cite{2,3,4}) one can find the anomaly induced
effective action \cite{2,3} (with accuracy up to conformally
invariant functional for the total effective action, see \cite{5} for a
review).  There is no consistent approach to calculate this conformally
invariant functional in closed form. However, one can find this
functional as some expansion of Schwinger--DeWitt type \cite{6}
keeping only the leading term.
Then, the effective action may be written in the following form
\cite{3,6}\footnote{Note that recently the effective action
(\ref{e4}) has been rederived in ref.\cite{7}.}
\be
\label{e4}
W=-{N \over 8\pi}\int d^2x \sqrt{-g}\,\left[
{1 \over 12}R{1 \over \Delta}R
-\nabla^\lambda \phi
\nabla_\lambda \phi {1 \over \Delta}R
+\phi R
+2\ln\mu^2 \nabla^\lambda \phi \nabla_\lambda \phi
\right],
\ee
where $\Delta$ is two--dimensional laplacian, $\mu^2$ is a dimensional
parameter. Here, the first term represents the Polyakov anomaly induced
action,
 the second and third terms give the dilaton dependent corrections to the
anomaly
induced action (these terms were found in ref.\cite{3} and first
discussed in connection with quantum gravity in  ref.
 \cite{4}, similar terms with slightly different coefficients were
derived in ref.\cite{2}). The last term (conformally invariant functional)
is found in ref.\cite{6}.
Note that it is possible to write the action (\ref{e3}) as some
non-linear $\sigma$-model with one loop quantum correction (\ref{e4}).
For this purpose, we rewrite (\ref{e4}) in a local form by introducing
auxilliary fields $u$ and $v$
\bea
\label{e4b}
W&=&-{N \over 8\pi}\int d^2x \sqrt{-g}\,\Bigl[
{1 \over 12}R v -\nabla^\lambda \phi
\nabla_\lambda \phi v +\phi R
+2\ln\mu^2 \nabla^\lambda \phi \nabla_\lambda \phi \nn
&& +\nabla^\lambda u \nabla_\lambda v + uR \Bigr]\ .
\eea
Then in the conformal gauge $g_{\mu\nu}=\e^{2\rho}\bar g_{\mu\nu}$,
we find
\be
\label{nlsa}
S_{red}+W=\int d^2x \sqrt{-\bar g}\left[{1 \over 2}G_{ij}(X)\bar g^{\mu\nu}
\partial_\mu X^i\partial_\nu X^j + \bar R\Phi(X) + T(X) \right],
\ee
where
\bea
\label{nlsb}
X^i&=&\{\phi,\rho,u,v,\chi_a\}\ ,\nn
\Phi(X)&=&-{1 \over 16\pi G}\e^{-2\phi}-{N \over 8\pi}\left({v \over  12}
+ \phi + v \right)\ ,\nn
T(X)&=&-{1 \over 16\pi G}\left(-2\Lambda\e^{2\rho-2\phi}
+ \e^{2\rho}\right)\ , \nn
G_{ij}&=&\left(\begin{array}{cccc|c}
-{\e^{-2\phi} \over 4\pi G} + {N \over 4\pi}\left(v - 2 \ln\mu^2  \right)
& {\e^{-2\phi} \over 4\pi G} - {N \over 4\pi} & 0 & 0 & 0 \\
{\e^{-2\phi} \over 4\pi G} - {N \over 4\pi} & 0 & - {N \over 4\pi} & 0 & 0 \\
0 & - {N \over 4\pi} & 0 & -{N \over 8\pi} & 0 \\
0 & 0 & -{N \over 8\pi} & 0 & 0 \\
\hline 0 & 0 & 0 & 0 & \e^{-2\phi} \\ \end{array}\right)\ .
\eea

Working in the conformal gauge
\be
\label{e5}
g_{\pm\mp}=-{1 \over 2}\e^{2\rho}\ ,\ \
g_{\pm\pm}=0,
\ee
the equations of motion may be obtained by the variation of
$\Gamma=S_{red}+W$ with respect to
$g^{\pm\pm}$, $g^{\pm\mp}$ and $\phi$
\bea
\label{e6}
0&=&{\e^{-2\phi} \over 4G}\left(
2\partial_r \rho\partial_r\phi + \left(\partial_r\phi\right)^2
-\partial_r^2\phi\right) \nn
&& -{N \over 12}\left( \partial_r^2 \rho - (\partial_r\rho)^2 \right)
- {N \over 2} \left(\rho+{1 \over 2}\right) (\partial_r\phi)^2 \nn
&& -{N \over 4}\left( 2 \partial_r \rho \partial_r \phi
- \partial_r^2 \phi \right) -{N \over 4}\ln \mu^2 (\partial_r\phi)^2
+ N t_0 \\
\label{e7}
0&=&{\e^{-2\phi} \over 8G}\left(2\partial_r^2 \phi
-4 (\partial_r\phi)^2 - 2\Lambda \e^{2\rho} +2 \e^{2\rho+2\phi}\right) \nn
&& +{N \over 12}\partial_r^2 \rho +{N \over 4}(\partial_r \phi)^2
-{N \over 4}\partial_r^2\phi \\
\label{e8}
0&=& -{\e^{-2\phi} \over 4G}\left(-\partial_r^2\phi
+(\partial_r\phi)^2
+\partial_r^2 \rho+ \Lambda \e^{2\rho}\right) \nn
&& + {N \over 4} \left\{2\partial_r(\rho \partial_r\phi)
+\partial_r^2\rho \right\} + {N \over 2}\ln \mu^2
\partial_r^2\phi \ .
\eea
Here, $t_0$ is a constant which is determined by the initial  conditions.
Below we are interested in the static solution that is why we replace
$\partial_\pm \rightarrow \pm{1 \over 2}\partial_r$
where $r$ is radial coordinate.

Combining (\ref{e6}) and (\ref{e7}) we get
\bea
\label{e9}
0&=&{\e^{-2\phi} \over 4G}\left(- \left(\partial_r\phi\right)^2
+ 2\partial_r \rho \partial_r \phi - \Lambda \e^{2\rho}
+ \e^{2\rho+2\phi}\right) \nn
&& +{N \over 12}(\partial_r\rho)^2 - {N \over 2} \rho  (\partial_r\phi)^2
 -{N \over 2} \partial_r \rho \partial_r \phi
 -{N \over 4}\ln \mu^2 (\partial_r\phi)^2+ N t_0 \ .
\eea
This equation may be used to determine $t_0$ from the initial  condition,
it decouples from the other two equations. Hence, Eq.(\ref{e9}) is not
necessary in subsequent analysis.

Below, we consider a subclass of metrics (\ref{e2}) of the following type
\be
\label{e10}
ds^2=-\e^{2\rho}dt^2 + dl^2 + \e^{-2\phi}d\Omega^2 ,
\ee
where $\rho=\rho(l)$, $\phi=\phi(l)$ and $l$ is the proper distance. Note
that to study exact solutions for matter in metrics of such sort, one
can use the methods developed in ref.\cite{19}.

In order to come to metrics of the form (\ref{e10}), it is convenient  to
change the radial coordinate $r$ to $l$ by
\be
\label{e11}
dl=\e^\rho dr\ .
\ee
Then we obtain
\be
\label{e12}
\partial_r = \e^\rho\partial_l\ ,\ \
\partial_r^2 = \e^{2\rho}\left(\partial_l^2
+ \partial_l\rho\partial_l \right)\ .
\ee
Eqs. (\ref{e7}) and (\ref{e8}) may be rewritten as follows
\bea
\label{e13}
0&=&\left({\e^{-2\phi} \over G}-N\right)\partial_l^2 \phi
+{N \over 3}\partial_l^2 \rho
+\left(-{2\e^{-2\phi} \over G}+N\right)(\partial_l\phi)^2 \nn
&& +\left({\e^{-2\phi} \over G}-N\right)\partial_l\rho \partial_l\phi
+ {1 \over G}\left( - \Lambda \e^{-2\phi} + 1 \right)
+{N \over 3}(\partial_l \rho)^2, \\
\label{e14}
0&=& -\left({\e^{-2\phi} \over G}-N\right)\partial_l^2 \rho
+ \left\{{\e^{-2\phi} \over G}
+ N \left(2\rho + 2\ln \mu^2\right) \right\}\partial_l^2 \phi
- {\e^{-2\phi} \over G} (\partial_l\phi)^2 \nn
&& + \left\{{\e^{-2\phi} \over G}
+ N \left(2\rho + 2 + 2\ln \mu^2\right)\right\}
 \partial_l\rho \partial_l\phi \nn
&& -\left({\e^{-2\phi} \over G}-N\right)(\partial_l\rho)^2
- {\e^{-2\phi} \over G}\Lambda \ .
\eea
Our next task will be the study of Eqs. (\ref{e13}), (\ref{e14}) for  the
wormhole metric (\ref{e10}), where the usual notations are
$f(l)=\exp (2\rho)$ (redshift function) and $r(l)=\exp (-\phi)$ (shape
function).
Actually, in our notations $r(l)$ is always positive, this function
gives the wormhole throat.

First, we consider purely induced gravity, i.e. $N\to\infty$ case. Then
Einstein action can be dropped away.
For this case, the field equations have the form
\bea
\label{e13b}
0&=&-\partial_l^2 \phi
+{1 \over 3}\partial_l^2 \rho
+(\partial_l\phi)^2
-\partial_l\rho \partial_l\phi+{1 \over 3}(\partial_l \rho)^2,\\
\label{e14b}
0&=& \partial_l^2 \rho
+ \left(2\rho + a\right) \partial_l^2 \phi
+\left(2\rho + 2 + a\right) \partial_l\rho \partial_l\phi
+(\partial_l\rho)^2 \ .
\eea
where $a=2\ln \mu^2$. These equations admit the next integrals of  motion
\bea
\label{e16}
I_1&=&\e^\rho\,\left(\rho'+(2\rho+ a)\phi'\right) \\
I_2&=&\e^{2\rho}\left((\rho')^2 - 3(2\rho + a)(\phi')^2
- 6 \rho'\phi'\right),
\eea
here $\ '\equiv \partial_l$. In our study we take the following initial
conditions:
\be
\label{e15}
f'(0)=r'(0)=0
\ee
For the conditions (\ref{e15}) we have $I_0=I_1=0$, we get the trivial
solution
\be
\label{e17}
\rho(l), \phi(l), \quad
r(l), f(l) = \mbox{const}.
\ee
This indicates the possibility of inducing wormholes with finite
throat.

For the case $f'_0=r'_0=0$ from equations (\ref{e13})--(\ref{e14}) in
point $l=0$ we have
\bea
\label{e13c}
0&=&\left({\e^{-2\phi_0} \over G}-N\right)\phi''_0
+{N \over 3}\rho''_0
+ {1 \over G}\left( - \Lambda \e^{-2\phi_0} + 1 \right), \\
\label{e14c}
0&=& -\left({\e^{-2\phi_0} \over G}-N\right)\rho''_0
+ \left\{{\e^{-2\phi_0} \over G}
+N \left(2\rho_0 + a\right)\right\}\phi''_0
- {\e^{-2\phi_0} \over G}\Lambda \ .
\eea
Let us consider the case when $\Lambda=0$ and $r(l)$, $f(l)$ are
non-decreasing functions near point $l=0$, then we have the next
restrictions for $\rho_0$ and $\phi_0$
\bea
&& \phi_0\le  -\frac{1}{2}\ln(GN)\ ,\nn
&& -{\e^{-2\phi_0} \over NG}-a
-\frac{3}{N^2}\left(N-{\e^{-2\phi_0} \over G}\right)^2<2\rho_0\le
-{\e^{-2\phi_0} \over NG}-a.
\eea
In other words, for the case $\Lambda=0$ the throat radius is
\be
r_0\ge\sqrt{GN}\ .
\ee

Let us consider equations (\ref{e13})--(\ref{e14}) for the case when
$r(l)=r_0=\mbox{const}$. We get two solutions
\begin{enumerate}
\item
$\Lambda\ne0$,
\quad
$|2GN\Lambda/3-2|\ge \sqrt{3}$,

$r_0{}^2=\left(2GN\Lambda/3+1\pm\sqrt{(2GN\Lambda/3-2)^2
-3}\, \right)/(2\Lambda)
$,

$
f(l)=\left\{
\begin{array}{ll}
(c_1\cosh\sqrt{k}\,l+c_2\sinh\sqrt{k}\,l)^2, & k>0 \\
(c_1+c_2l)^2, & k=0 \\
(c_1\cos\sqrt{-k}\,l+c_2\sin\sqrt{-k}\,l)^2, & k<0, {\rm with}
\end{array}
\right.
$

$
k= (\Lambda r_0{}^2-1)/G,
$

\item
$\Lambda=0$,

$r_0=\sqrt{GN}$,
\quad
$f(l)=\left(
\sqrt{f_0}\cos\sqrt{3/GN}\,l+
(r_0f'_0/2\sqrt{3f_0})\sin\sqrt{3/GN}\,l
\right)^2.$
\end{enumerate}
Hence we found the wormhole (as a quantum
effect induced object) with a constant throat radius and increasing
or decreasing redshift function $f(l)$. This wormhole connects two spatial
regions of spacetime.
This can be viewed as  a kind of infinitely long wormhole.

>From another side in some above cases we get oscillating behaviour
for $f(l)$. This fact indicates to some instability of wormhole
 configuration. Hence one can expect that kind of topological
phase transition may occur and transform the wormhole to
some another object, like black hole.

Let us discuss now the case when the classical gravity action is included,
i.e. eqs.(\ref{e13}), (\ref{e14}) but $r$ is not constant. In this case
we perform a numerical study of the  equations of motion.
Considering for simplicity $\Lambda=0$ and $\ln\mu^2=1$, we present in
 figures 1. and 2. the typical result of our numerical study.
The redshift function
$f$ increases as well as $r(l)$. It is interesting that
qualitatively $r(l)$ and $f(l)$ behave in the same way as
in ref.\cite{17}. In particulary, the redshift function
does not approach to constant value at large $l$.,
hence the whole metric is not asymptoticaly flat.

\epsffile[0 320 500 550]{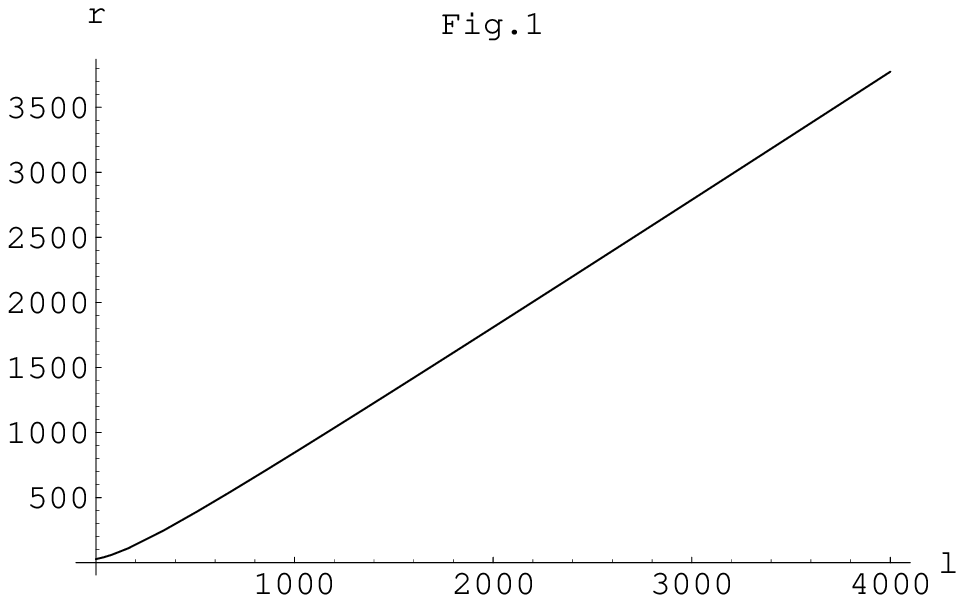}

\epsffile[0 320 500 550]{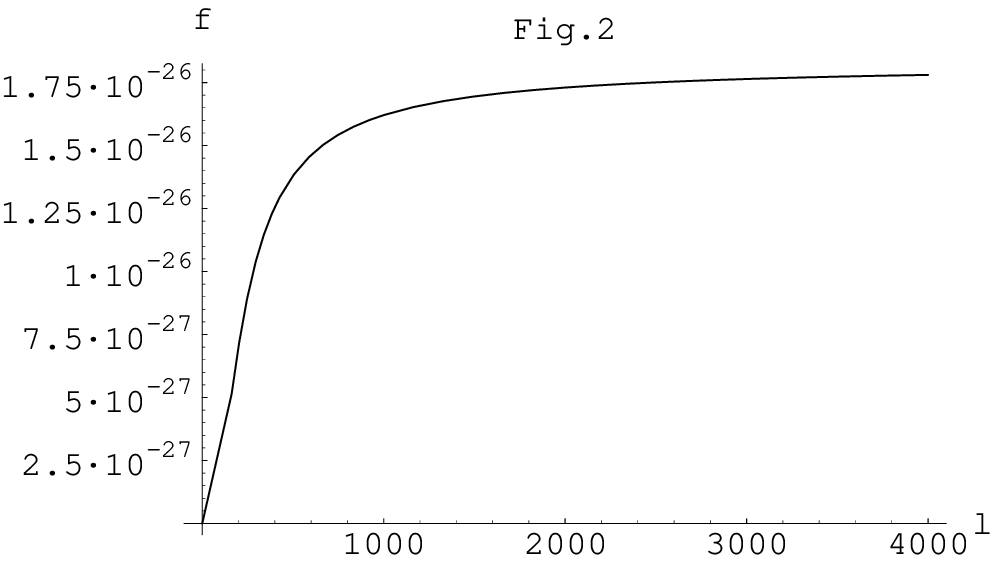}
\noindent
In these figures  $r_0=27.1828$, $f_0=2.1405\cdot 10^{-31}$,
$r'_0=f'_0=0$ and $N=100$.

Hence, we showed that even for the whole system, i.e. classical gravity
plus the quantum effective action due to matter, we can expect the
appearance of primordial (induced) wormholes in the early Universe. The
wormhole which appears as result of quantum fluctuations connects two
asymptotically (flat) regions of the early Universe. It may be with constant
or increasing throat radius. The redshift function maybe increasing or
oscillating. It would be very interesting to generalize above study
for all types of quantum fields. Then one could investigate the connection
between content of GUTs and inducing of quantum wormholes. For example,
it may happen that SUSY GUTs better support inducing of wormholes
as they include more fields then their non-SUSY versions.

\

\noindent
{\bf Acknowledgments}.
SDO would like to thank S. Hawking and E. Mottola for useful  discussion.
O.O. was partially supported by a CONACYT Grant 28454-E.
KEO was partially supported by a RFBR.

\end{document}